\documentclass[preprint,showpacs,preprintnumbers,amsmath,amssymb]{revtex4}

\usepackage{graphicx}
\usepackage{dcolumn}
\usepackage{bm}

\begin{document}

\newcommand*{\PKU}{School of Physics and State Key Laboratory of Nuclear Physics and
Technology, Peking University, Beijing 100871, China}\affiliation{\PKU}
\newcommand*{\chep}{Center for High Energy Physics, Peking University, Beijing 100871, China}\affiliation{\chep}

\title{Tetramixing of vector and pseudoscalar mesons: A source of intrinsic quarks}

\author{Tao Peng}\affiliation{\PKU}
\author{Bo-Qiang Ma}\email{mabq@pku.edu.cn}\affiliation{\PKU}\affiliation{\chep}

\date{\today}

\begin{abstract}
The tetramixing of pseudoscalar mesons
$\pi$-$\eta$-$\eta'$-$\eta_c$ and vector mesons
$\omega$-$\rho$-$\phi$-$J/\psi$ are studied in the light-cone
constituent quark model, and such mixing of four mesons provides a
natural source for the intrinsic charm $c\bar{c}$ components of
light mesons. By mixing with the light mesons, the charmonium
states $J/\psi$ and $\eta_c$ could decay into light mesons more
naturally, without introducing gluons or a virtual photon as
intermediate states. Thus, the introduction of light quark components
into $J/\psi$ is helpful to reproduce the new experimental data of
$J/\psi$ decays. The mixing matrices and the $Q^2$ behaviors of the
transition form factors are also calculated and compared with
experimental data.
\end{abstract}


\pacs{12.39.Ki, 13.20.Gd, 13.40.Gp, 14.40.Be}%

\maketitle

The mixing of mesons has been widely investigated since the 1960s,
when the concept of a mixing state of the $\rho$-$\omega$ mesons was
proposed~\cite{Glashow61} by considering that the electromagnetic
interaction does not conserve isospin. Later, the $\omega$-$\phi$
mixing and $\eta$-$\eta'$ mixing were introduced~\cite{Sakurai62} to
explain the deviation of the meson mass from the Gell-Mann-Okubo
mass formula~\cite{Gell-Mann61,Gell-Mann62}. It was also pointed out
that the difference between the $u$ and $d$ quark masses introduces
the $\pi$-$\eta$ mixing~\cite{Gross79}. Then, the trimixing of
$\pi$-$\eta$-$\eta'$~\cite{Gusbin81,Qian09} and
$\rho$-$\omega$-$\phi$~\cite{Benayoun01,Qian09} were proposed, and
their effects were studied in different methods. On the other hand,
the $c\bar{c}$ contribution to the $\eta$ and $\eta'$ mesons was
considered~\cite{Harari75}, and the trimixing
$\eta$-$\eta'$-$\eta_c$ was studied~\cite{Fritsch77}. As a further
extension in this paper, we try to combine the above two types of
trimixing by considering the tetramixing of pseudoscalar mesons
$\pi$-$\eta$-$\eta'$-$\eta_c$. The mixing of gluon component $gg$
and $\eta$-$\eta'$ were also studied~\cite{Donato11,Ke10}. As the
mixing of $\eta$ and $\eta'$ is still not completely clear right
now, we think that the charm and gluon components may both be
possible to mix with these mesons, and it is worthwhile to study both
of them carefully.

The recent CLEO experiment~\cite{CLEO09} of the charmonium decays
$J/\psi \rightarrow \gamma \pi$,
 $\gamma \eta$, and $\gamma \eta '$ also motivates us to extend our
tetramixing to the vector mesons $\omega$-$\rho$-$\phi$-$J/\psi$.
According to the pure valence
$c\bar{c}$ structure of charmonia in the naive quark model, these
decay modes of charmonium $\psi(nS)$ must happen via the
annihilation of the heavy quark constituents into gluons or a
virtual photon~\cite{Brodsky97,CLEO09}, because of the
Okubo-Zweig-Iizuka rule, which postulates a suppression of
transitions between hadrons without valence quarks in
common~\cite{Brodsky97}. Moreover, the mechanisms of these decays
are not completely clear yet, and there are various ways to describe
them, such as $\psi(nS)\rightarrow \gamma gg\rightarrow \gamma  P$,
$\psi(nS)\rightarrow ggg \rightarrow q\bar{q}\gamma$,
$\psi(nS)\rightarrow \gamma^*\rightarrow q\bar{q}\gamma$, and
so on~\cite{CLEO09}. However, with the model of
$\omega$-$\rho$-$\phi$-$J/\psi$ mixing, the above-mentioned decays
of $J/\psi$ could occur more naturally through the direct transition
from its light quark components to light mesons such as $\pi$,
$\eta$, or $\eta'$ without introducing intermediate gluons or a
virtual photon, and the $c\bar{c}$ components of these light
pseudoscalar mesons also allow\ $J/\psi$ to decay to them. The
mixing of $\omega$-$\rho$-$\phi$-$J/\psi$ is thus helpful to
reproduce the new experimental data of $J/\psi$ decays.

For the light vector mesons $\omega$, $\rho$, and $\phi$, the
existence of $c\bar{c}$ states in them may be interpreted as a
support to the theory of intrinsic charm~\cite{Brodsky80} in these
mesons. Different from the extrinsic quarks, which are generated on a
short time scale in a reaction process with large momentum
transfers, the intrinsic quarks are intrinsic nonperturbatively to
the hadron wave function and exist over a time scale independent of
any probe momentum~\cite{Brodsky80,Brodsky81}. The postulation of
intrinsic $c\bar{c}$ components in $\rho$ and $\pi$ offers a
possible solution of the ``$\rho \pi$ puzzle" by allowing direct
transitions between $J/\psi(\psi')$ and $\rho(\pi)$ through the
rearrangement of the valence and the intrinsic $c\bar{c}$ components
of $\rho(\pi)$~\cite{Brodsky97}. Now the tetramixing of
$\omega$-$\rho$-$\phi$-$J/\psi$ introduces the intrinsic $c\bar{c}$
components into all three light vector mesons $\omega$,
$\rho$, and $\phi$, and $J/\psi$ can decay to them in a similar way,
without annihilation of the quark constituents. This applies to the
pseudoscalar mesons $\pi$, $\eta$, and $\eta'$, too, as they mix with
the charmonium $\eta_c$ in our model. The intrinsic $c\bar{c}$
component in $\eta'$ was also studied in Refs.~\cite{Cao98,Huang07},
in which the intrinsic charm content of the $\eta'$ meson
$f_{\eta'}^c$ was evaluated, and we shall compare our result of
$f_{\eta'}^c$ with previous results in Refs.~\cite{Cao98,Huang07}
and other works at the end of this paper.

We adopt the light-cone constituent quark
model~\cite{Lepage80,Brodsky82,Brodsky98} to study the mixing of
mesons. The light-cone constituent quark model is a convenient and
effective model to treat the nonperturbative aspect of QCD, and the
mixing of mesons in this model has been
studied~\cite{Xiao05,Qian08}.

The mixing of pseudoscalar mesons and vector mesons could be described
by two SO(4) rotation matrices $M_v$ and $M_s$, respectively:
\begin{eqnarray}
\left(
  \begin{array}{c}
  \omega \\
  \rho \\
  \phi\\
  J/\psi\\
  \end{array}
\right) = M_v \left(
  \begin{array}{c}
  \omega_I \\
  \rho_I \\
  \phi_I \\
  J/\psi_I\\
  \end{array}
\right), ~~ \left(
  \begin{array}{c}
  \pi \\
  \eta \\
  \eta'\\
  \eta_c\\
  \end{array}
\right) = M_s \left(
  \begin{array}{c}
  \pi_I \\
  \eta_q \\
  \eta_s \\
  \eta_{c0}\\
  \end{array}
\right),
\label{eq:states}
\end{eqnarray}
in which the unmixed states are
\begin{eqnarray}
\begin{array}{cccc}
\omega_I = \frac{1}{\sqrt{2}}(u\bar{u}+d\bar{d})\varphi_{\omega_I},
& \rho_I = \frac{1}{\sqrt{2}}(u\bar{u}-d\bar{d})\varphi_{\rho_I},
& \phi_I = -s\bar{s}\varphi_{\phi_I},
& J/\psi_I = c\bar{c}\varphi_{J/\psi_I},\\
\pi_I = \frac{1}{\sqrt{2}}(u\bar{u}-d\bar{d})\varphi_{\pi_I},
& \eta_q = \frac{1}{\sqrt{2}}(u\bar{u}+d\bar{d})\varphi_{\eta_q},
& \eta_s = s\bar{s}\varphi_{\eta_s},
& \eta_{c0} = c\bar{c}\varphi_{\eta_{c0}},
\end{array}
\end{eqnarray}
where $\varphi$ is the momentum space wave function of the corresponding meson.

Since the rotation group $SO(4)=SO(3)\otimes SO(3)$, the $SO(4)$ mixing matrix $M$ can be  written as
\begin{equation}
M=R_+ R_-,
\end{equation}
where the matrices $R_+$ and $R_-$ are generated by the $SO(3)$
generators, and each of them could be parameterized by three
independent rotation angles as
\begin{eqnarray}
R_+(\theta_1,\theta_2,\theta_3) = \left( \begin{array}{cccc}
\cos \frac{\alpha}{2} & -\frac{\theta_1}{\alpha}\sin \frac{\alpha}{2} & -\frac{\theta_2}{\alpha}\sin \frac{\alpha}{2}
& -\frac{\theta_3}{\alpha}\sin \frac{\alpha}{2}\\
\frac{\theta_1}{\alpha}\sin \frac{\alpha}{2} & \cos \frac{\alpha}{2} & -\frac{\theta_3}{\alpha}\sin \frac{\alpha}{2}
& \frac{\theta_2}{\alpha}\sin \frac{\alpha}{2}\\
\frac{\theta_2}{\alpha}\sin \frac{\alpha}{2} & \frac{\theta_3}{\alpha}\sin \frac{\alpha}{2} & \cos \frac{\alpha}{2}
& -\frac{\theta_1}{\alpha}\sin \frac{\alpha}{2}\\
\frac{\theta_3}{\alpha}\sin \frac{\alpha}{2} & -\frac{\theta_2}{\alpha}\sin \frac{\alpha}{2} & \frac{\theta_1}{\alpha}\sin \frac{\alpha}{2} & \cos \frac{\alpha}{2}\\
\end{array}\right),
\label{eq:rp}
\end{eqnarray}

\begin{eqnarray}
R_-(\theta_4,\theta_5,\theta_6) = \left( \begin{array}{cccc}
\cos \frac{\beta}{2} & \frac{\theta_4}{\beta}\sin \frac{\beta}{2} & \frac{\theta_5}{\beta}\sin \frac{\beta}{2}
& \frac{\theta_6}{\beta}\sin \frac{\beta}{2}\\
-\frac{\theta_4}{\beta}\sin \frac{\beta}{2} & \cos \frac{\beta}{2} & -\frac{\theta_6}{\beta}\sin \frac{\beta}{2}
& \frac{\theta_5}{\beta}\sin \frac{\beta}{2}\\
-\frac{\theta_5}{\beta}\sin \frac{\beta}{2} & \frac{\theta_6}{\beta}\sin \frac{\beta}{2} & \cos \frac{\beta}{2}
& -\frac{\theta_4}{\beta}\sin \frac{\beta}{2}\\
-\frac{\theta_6}{\beta}\sin \frac{\beta}{2} & -\frac{\theta_5}{\beta}\sin \frac{\beta}{2}
& \frac{\theta_4}{\beta}\sin \frac{\beta}{2} & \cos \frac{\beta}{2}\\
\end{array}\right),
\label{eq:rm}
\end{eqnarray}
where
\begin{equation}
\alpha=\sqrt{\theta_1^2+\theta_2^2+\theta_3^2},~~\beta=\sqrt{\theta_4^2+\theta_5^2+\theta_6^2},
\end{equation}
and thus the mixing matrix $M$ is parameterized as six independent
rotation angles~($\theta_1$, $\theta_2$, ... , $\theta_6$). Our
detailed procedure of obtaining the matrix form of $R_+$ and
$R_-$~[Eqs.(4) and (5)] is given in Appendix~\ref{app:a}.

When referring to the mixing of specific types of mesons, $M$ stands for $M_v$ or $M_s$,
and the parameters change to ($\theta_1^v$, $\theta_2^v$, ...) or ($\theta_1^s$, $\theta_2^s$, ...) correspondingly.

During the numerical calculation, we also used a more compact form
of $M$ with eight real parameters under constraints, and the
detailed procedure is given in Appendix~\ref{app:b}.

The decay constants and transition form factors also mix as~\cite{Qian09}
\begin{equation}
\left(
\begin{array}{l}
 f_\omega  \\
 f_\rho  \\
 f_\phi \\
 f_{J/\psi}
\end{array}
\right) = M_v \left(
\begin{array}{l}
 f_{\omega_I} \\
 f_{\rho_I} \\
 f_{\phi_I}\\
 f_{{J/\psi}_I}
\end{array}
\right), ~~ \left(
\begin{array}{l}
 F_{\pi\rightarrow\gamma\gamma^*} (Q^2) \\
 F_{\eta\rightarrow\gamma\gamma^*} (Q^2)  \\
 F_{\eta'\rightarrow\gamma\gamma^*}(Q^2) \\
 F_{\eta_c \rightarrow \gamma \gamma^*} (Q^2)
\end{array}
\right) = M_s \left(
\begin{array}{l}
 F_{\pi_I\rightarrow\gamma\gamma^*}(Q^2)  \\
 F_{\eta_q\rightarrow\gamma\gamma^*}(Q^2)  \\
 F_{\eta_s\rightarrow\gamma\gamma^*}(Q^2)\\
 F_{\eta_{c0}\rightarrow\gamma\gamma^*}(Q^2)
\end{array}
\right),~~
\end{equation}
\begin{equation}
\left(
\begin{array}{l}
 F_{\omega \rightarrow\pi \gamma^* } (Q^2)\\
 F_{\omega \rightarrow\eta \gamma^* } (Q^2)\\
 F_{\eta' \rightarrow\omega \gamma^* }(Q^2) \\
 F_{\eta_c \rightarrow\omega \gamma^* }(Q^2) \\
 F_{\rho \rightarrow\pi \gamma^* }(Q^2) \\
 F_{\rho \rightarrow\eta \gamma^* } (Q^2)\\
 F_{\eta' \rightarrow\rho \gamma^* } (Q^2)\\
 F_{\eta_c \rightarrow\rho \gamma^* } (Q^2)\\
 F_{\phi \rightarrow\pi \gamma^* }(Q^2)\\
 F_{\phi \rightarrow\eta \gamma^* }(Q^2)\\
 F_{\phi \rightarrow\eta' \gamma^* }(Q^2)\\
 F_{\eta_c \rightarrow\phi \gamma^* }(Q^2)\\
 F_{J/\psi \rightarrow\pi \gamma^* } (Q^2)\\
 F_{J/\psi \rightarrow\eta \gamma^* } (Q^2)\\
 F_{J/\psi \rightarrow\eta' \gamma^* } (Q^2)\\
 F_{J/\psi \rightarrow\eta_c \gamma^* } (Q^2)
\end{array}
\right) = ( M_v\otimes M_s )
 \left(
\begin{array}{l}
 F_{\omega_I\rightarrow\pi_I \gamma^* }(Q^2) \\
 F_{\omega_I\rightarrow\eta_q \gamma^* }(Q^2)\\
 0\\
 0\\
 F_{\rho_I\rightarrow\pi_I \gamma^* } (Q^2)\\
 F_{\rho_I\rightarrow\eta_q \gamma^* }(Q^2) \\
 0\\
 0\\
 0\\
 0 \\
 F_{\phi_I\rightarrow\eta_q \gamma^* }(Q^2)\\
 0\\
 0\\
 0\\
 0\\
 F_{J/\psi_I \rightarrow\eta_{c0} \gamma^* } (Q^2)
\end{array}
\right).
\end{equation}
The above decay constants and transition form factors are defined
as~\cite{Qian08,Choi97}
\begin{equation}
\langle 0| j_\mu |V(p,S_z)\rangle=M_V f_V \epsilon_\mu(S_z),
\end{equation}
\begin{equation}
\langle \gamma(p-q)|J^\mu |P(p,\lambda)\rangle
=ie^2F_{P\rightarrow \gamma\gamma^*}(Q^2)
        \varepsilon^{\mu\nu\rho\sigma}p_\nu\epsilon_\rho(p-q,\lambda)
        q_\sigma,
\end{equation}
\begin{equation}
\langle P(p')|J^\mu |V(p,\lambda)\rangle
 = ieF_{V\rightarrow P\gamma}(Q^2)
        \varepsilon^{\mu\nu\rho\sigma}\epsilon_\nu(p,\lambda) p'_\rho
        p_\sigma.
\end{equation}
To calculate them, we use the light-cone quark model with the Fock
state expansions of the unmixed mesons [the right-hand side of
Eq.~(\ref{eq:states})]:
\begin{eqnarray}
|M\rangle &=& \sum |q\bar{q}\rangle \psi_{q\bar{q}}
        + \sum
        |q\bar{q}g\rangle \psi_{q\bar{q}g} + \cdots,
\end{eqnarray}
and to simplify the problem, we adopt the lowest order of the above expansions, which takes
only the quark-antiquark valence states of the unmixed mesons into consideration.

The wave function of an unmixed meson in the light-cone formalism is~\cite{Brodsky82,Huang94}
\begin{equation}
|M (P^+, \mathbf{P}_\perp, S_z) \rangle
   =  \int \frac{\mathrm{d} x \mathrm{d}^2
        \mathbf{k}_{\perp}}{\sqrt{x(1-x)}16\pi^3}
        \varphi(x,\mathbf{k}_{\perp})
        \chi_M^{S_z}(x,\mathbf{k}_{\perp},\lambda_1,\lambda_2),
\end{equation}
where $\varphi$ is the momentum space wave function, described by the Brodsky-Huang-Lepage prescription~\cite{Brodsky82,Huang94}:
\begin{equation}
\varphi(x,\mathbf{k}_{\perp}) =
\varphi_{\mathrm{BHL}}(x,\mathbf{k}_{\perp})
=A \exp
\left[-\frac{1}{8\beta^2}\left(\frac{m_1^2+\mathbf{k}_\perp^2}{x}
+\frac{m_2^2+\mathbf{k}_\perp^2}{1-x}\right)\right],
\end{equation}
($A$ and $\beta$ are the parameters of the meson, and $m_1$ and
$m_2$ are masses of the constituent quarks), and $\chi_M^{S_z}(x,
\mathbf{k}_{\perp}, \lambda_1, \lambda_2)$ is the light-cone spin
wave function, which is related to the instant-form spin wave
function by the Melosh-Wigner
rotation~\cite{Kondratyuk80,Melosh74,Ma93}
\begin{equation}
\chi_i^\uparrow(T) = w_i[(k_i^+ +m_i)\chi_i^\uparrow(F)-k_i^R
\chi_i^\downarrow(F)];
~~\chi_i^\downarrow(T)= w_i[(k_i^+ +m_i)\chi_i^\downarrow(F)+k_i^L
\chi_i^\uparrow(F)],
\end{equation}
where $w_i=1/\sqrt{2k_i^+ (k^0+m_i)}$, $k^{R,L}=k^1\pm k^2$,
$k^+=k^0+k^3=x \mathcal{M}$, and
$\mathcal{M}=\sqrt{(\mathbf{k_\perp}^2+m_1^2)/x+(\mathbf{k_\perp}^2+m_2^2)/(1-x)}$.
The Melosh-Wigner rotation is an important ingredient of the
light-cone quark model and plays an essential role in explaining the
``proton spin puzzle" ~\cite{Ma93,Ma96}. The detailed formulas for
calculating the decay constants and transition form factors of
mesons were listed in Ref.~\cite{Qian08}, and the examples of
applying them to set meson parameters and to calculate the decay
constants and transition form factors numerically can be found in
Ref.~\cite{Xiao}.

The values of the meson parameters $A$, $\beta$, $m_1$, and $m_2$ and the parameters of the mixing matrices
($a_v$, $b_v$, ...) and ($a_s$, $b_s$, ...) can be chosen by fitting the light-cone constituent
quark model results of the meson decay constants and transition form factors (at $Q^2=0$) to experimental data.
The $Q^2\rightarrow\infty$ limiting behavior of $Q^2
F_{P\rightarrow\gamma\gamma^*}$ is also considered as a constraint
to set the parameters~\cite{Lepage80,Cao99}:
\begin{equation}
\lim_{Q^2\rightarrow\infty}Q^2F_{P\rightarrow\gamma\gamma^*}(Q^2) =
2c_P f_P = \frac{2 c_P^2}{4\pi^2 F_{P\rightarrow\gamma\gamma^*}(0)},
\label{eq:limitting}
\end{equation}
where $c_P =(c_{\pi_I}$, $c_{\eta_q}$, $c_{\eta_s})=(1$, $5/3$, $\sqrt{2}/3).$ With the similar method as Ref.~\cite{Cao98}, we also obtain
\(c_{\eta_{c0}}= 4\sqrt{2}/3\). All these requirements are taken as constraints to determine the parameters of mesons and parameters of the mixing matrices. During our calculation, we first use the decay constants and the radii of $\pi^+$ and $K^+$ as the constraints to locate the values of $A_\pi$, $\beta_\pi$, $m_u$, and $m_s$, assuming that the wave function parameters of $\pi^\pm$ are the same as those of $\pi_I$, and the constituent quark mass $m_d \approx m_u$ (their difference could be ignored compared with $m_c$)~\cite{Qian09}. The other parameters are then determined under the left constraints.

Our numerical calculation gives the mixing matrices of vector and pseudoscalar mesons:
\begin{eqnarray}
M_v = \left( \begin{array}{cccc}
0.9886 &  -0.0122  & -0.1429 &   0.0076\\
0.0299  &  0.9910  &  0.1221  & -0.0025\\
0.1400  & -0.1250  &  0.9808  &  0.0258\\
-0.0111  &  0.0058  & -0.0239 &   0.9986
\end{array}\right),
\end{eqnarray}
\begin{eqnarray}
M_s = \left( \begin{array}{cccc}
0.9895  &  0.0552  & -0.1119  &  0.0342\\
-0.1082  &  0.8175  & -0.5614  & -0.0259\\
0.0590   & 0.5696  &  0.8160  &  0.0452\\
-0.0395  & -0.0065 &  -0.0478   & 0.9960
\end{array}\right).
\end{eqnarray}
We see that some of the entries of the mixing matrices are small;
for example, one of the entries in the first row of $M_v$ is 0.0076.
But this nonzero entry means
 a charm component in the $\omega$ meson, which allows $\eta_c$ to decay to $\omega$
directly in our model. Other entries of the mixing matrices have the same meaning,
and it is such entries that are helpful to reproduce the experimental decay data
of $J/\psi$ and other meson decays.

The results of fitting light-cone constituent quark model results to
experimental data are shown in Table~\ref{fitting results}. The
fourth column contains the results of tetra-mixing model
$\pi$-$\eta$-$\eta'$ and $\rho$-$\omega$-$\phi$, while $J/\psi$ and
$\eta_c$ do not mix with other meson states, with the values of
their parameters~(MeV) set as $A_{J/\psi}=31.1660$,
$\beta_{J/\psi}=0.9777$, $A_{\eta_{c}}=125.7935$, and
$\beta_{\eta_{c}}=0.7524$ to fit the experimental data. The most
apparent differences between tetramixing and trimixing results are
in the last four rows, which show that the trimixing formalism do
not explain the nonzero decay width of $J/\psi \rightarrow \pi$,
$\eta$, and $\eta'$, while the tetramixing formalism can well
reproduce these experimental decay data. The parameters of the
mesons and the mixing matrices determined during the fitting process
are listed in Tables~\ref{meson parameters} and Table~\ref{matrix
parameters}.

\newcommand{\tabincell}[2]{\begin{tabular}{@{}#1@{}}#2\end{tabular}}
\begin{table}
\caption{\label{fitting results}Experimental data and the light-cone
constituent quark model fitting results of the meson decay constants
and transition form factors. The experimental data~(unmarked) are
from Ref.~\cite{PDG10}, and the experimental data~(marked with
daggers) are from Ref.~\cite{CLEO09}. The data in the fourth column
~(unmarked) are from Ref.~\cite{Qian09}, and the data~(marked with stars) are
calculated assuming that $J/\psi$ and $\eta_c$ do not mix.}
\begin{tabular}{cccc}
\hline
\hline
\tabincell{c}{Decay Constants \\ or Form Factors}  & \tabincell{c}{Experimental\\ Data~(GeV)} & \tabincell{c}{Theoretical Fitting\\ of tetramixing~(GeV)}
& \tabincell{c}{Theoretical Fitting\\ of trimixing~(GeV)}\\
\hline
$F_{\pi\rightarrow\gamma\gamma^*}(0)$ & 0.2744 $\pm$ 0.0082  & 0.2909 & 0.279\\
$F_{\eta \rightarrow\gamma\gamma^*}(0)$ & 0.2726 $\pm$ 0.0074 &  0.2891 & 0.277\\
$F_{\eta' \rightarrow\gamma\gamma^*}(0)$ & 0.3423 $\pm$ 0.0101 &  0.3187 & 0.334\\
$F_{\eta_c \rightarrow\gamma\gamma^*}(0)$& 0.0806 $\pm$ 0.0004 & 0.0568 & $0.0810^*$\\
\hline
$f_\omega(\omega\rightarrow e^+ e^-)$ & 0.0466 $\pm$ 0.0005 &  0.0502 & 0.04556\\
$f_{\rho}(\rho\rightarrow e^+ e^-)$ & 0.1549 $\pm$ 0.0009 &  0.1815 & 0.1603\\
$f_\phi(\phi\rightarrow e^+ e^-)$ & 0.0758 $\pm$ 0.0005 &  0.0729 & 0.075\\
$f_{J/\psi}(J/\psi \rightarrow e^+ e^-)$ & 0.2768 $\pm$ 0.0044  &  0.2734 & $0.2749^*$\\
\hline
$F_{\omega\rightarrow\pi\gamma^*}(0)$ & 2.2978 $\pm$ 0.0403 &  2.4639 & 2.382\\
$F_{\omega\rightarrow\eta \gamma^*}(0)$ & 0.4494 $\pm$ 0.0197 & 0.4285  & 0.454\\
$F_{\eta'\rightarrow\omega\gamma^*}(0)$ & 0.4260 $\pm$ 0.0355 &  0.4528 & 0.461\\
$F_{\eta_c\rightarrow\omega\gamma^*}(0)$ & ? &  -0.0895 & 0\\
\hline
$F_{\rho\rightarrow\pi\gamma^*}(0)$ & 0.8237 $\pm$ 0.0549 &  0.9207 & 0.84\\
$F_{\rho\rightarrow\eta\gamma^*}(0)$ & 1.5687 $\pm$ 0.0525 &  1.6124 & 1.50\\
$F_{\eta'\rightarrow\rho\gamma^*}(0)$ & 1.3175 $\pm$ 0.0327 &  1.3818 & 1.39\\
$F_{\eta_c\rightarrow\rho\gamma^*}(0)$& ? & -0.0537 & 0\\
\hline
$F_{\phi \rightarrow\pi \gamma^* }(0)$ & 0.1331 $\pm$ 0.0032 &  0.1301 & 0.132\\
$F_{\phi \rightarrow\eta \gamma^* }(0)$ & -0.6937 $\pm$ 0.0071 &  -0.7106 & -0.677\\
$F_{\phi \rightarrow\eta' \gamma^* }(0)$ & 0.7153 $\pm$ 0.0125 &  0.7261 & 0.727\\
$F_{\eta_c \rightarrow\phi \gamma^* }(0)$ & ? & -0.0404 & 0\\
\hline
$F_{J/\psi \rightarrow\pi \gamma^* } (0)$ & 0.0006 $\pm$ $0.000~03^{\dagger}$ & 0.0006 & 0\\
$F_{J/\psi \rightarrow\eta \gamma^* } (0)$ & 0.0035 $\pm$ $0.000~07^{\dagger}$ &0.0035 & 0\\
$F_{J/\psi \rightarrow\eta' \gamma^* } (0)$ & 0.0085 $\pm$ $0.0002^{\dagger}$ &0.0083 & 0\\
$F_{J/\psi \rightarrow\eta_c \gamma^* } (0)$ & 0.6583 $\pm$ $0.0787$&0.5991 & $0.6545^*$\\
\hline
\hline
\end{tabular}
\end{table}

\begin{table}
\caption{\label{meson parameters}The meson parameters $A$ and
$\beta$~(GeV), and the masses~(GeV) of constituent quarks determined
from the fitting process.}
\begin{tabular}{cccccccccc}
\hline
\hline
$A_{\omega_I}$ & $A_{\rho_I}$ & $A_{\phi_I}$ & $A_{J/\psi_I}$ & $A_{\pi_I}$ & $A_{\eta_q}$ & $A_{\eta_s}$ & $A_{\eta_{c0}}$ & $m_{u(d)}$& $m_s$ \\
41.4712 & 38.1430 & 63.1638 & 31.1724 & 47.3635 & 38.7860 & 95.4496 & 125.8099 & 0.198 & 0.556\\
\hline
$\beta_{\omega_I}$ & $\beta_{\rho_I}$ & $\beta_{\phi_I}$ & $\beta_{J/\psi_I}$ & $\beta_{\pi_I}$ & $\beta_{\eta_q}$ & $\beta_{\eta_s}$ & $\beta_{\eta_{c0}}$&$m_c$ &\\
0.4319 & 0.4318 & 0.4757 & 0.9781 & 0.4112 & 0.4887 & 0.4887 & 0.7373 & 1.270&\\
\hline
\hline
\end{tabular}
\end{table}

\begin{table}
\caption{\label{matrix parameters}Parameters of the mixing matrices
$M_v$ and $M_s$ determined from the fitting process.}
\begin{tabular}{cccccc}
\hline
\hline
$\theta_1^v$ & $\theta_2^v$ & $\theta_3^v$ & $\theta_4^v$ & $\theta_5^v$ & $\theta_6^v$\\
$-0.2181^\circ$ & $7.9190^\circ$ & $-7.6665^\circ$ & $-2.6511^\circ$ & $8.4010^\circ$ &  $-6.5877^\circ$\\
\hline
$\theta_1^s$ & $\theta_2^s$ & $\theta_3^s$ & $\theta_4^s$ & $\theta_5^s$ & $\theta_6^s$\\
$-7.8710^\circ$ & $4.6505^\circ$ & $32.4605^\circ$ & $2.1461^\circ$ & $5.8085^\circ$ &  $36.7524^\circ$\\
\hline
\hline
\end{tabular}
\end{table}

The $Q^2$ behaviors of the form factors of $\pi$, $\eta$, and
$\eta'$ are shown in Figs. 1-3, and we see that they are generally
in agreement with the experimental data. The $Q^2$ behavior of the
form factor of $\eta_c$ is shown in Fig. 4, and, by comparing with
theoretical data from another model, the calculated curve fits well
in most of the lower $Q^2$ region. We can also obtain the $Q^2$
behavior of the transition form factors in the timelike region,
either by making the substitution $q_\perp\rightarrow iq_\perp
$~\cite{Choi01} or by parameterizing the transition form factors as
explicit functions of $q^2$ in the spacelike region and then
extending them through analytic continuum to the timelike
region~\cite{Wang10}. The results are shown in Figs.~5-10, among
which Figs.~5 and Fig.~6 are compared with the experimental data,
while Figs.~7-10 could be considered as our predictions of the $Q^2$
behaviors of $J/\psi$ transition form
factors.

\begin{figure}[h]
\centering
\includegraphics[0,0][300,250]{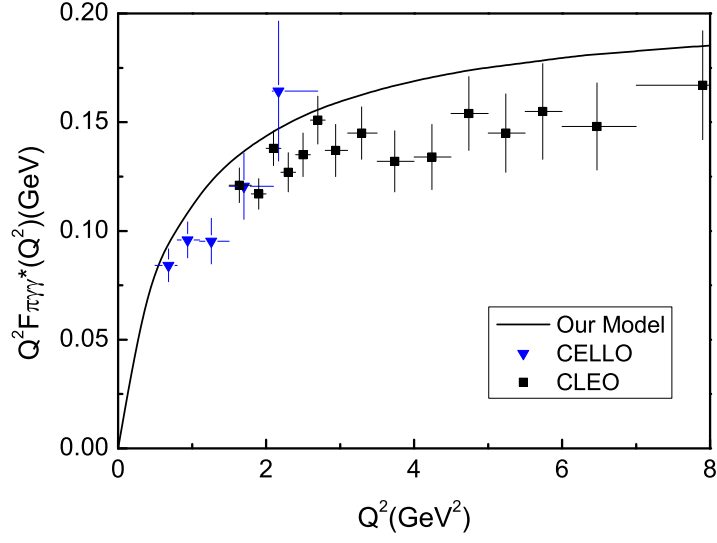}
\caption{The $Q^2$ behavior of the form factor $Q^2
F_{\pi\rightarrow\gamma\gamma^*}(Q^2)$ compared with experimental
data~\cite{Behrend91,Gronberg98}.}
\end{figure}

\begin{figure}[h]
\centering
\includegraphics[0,0][300,250]{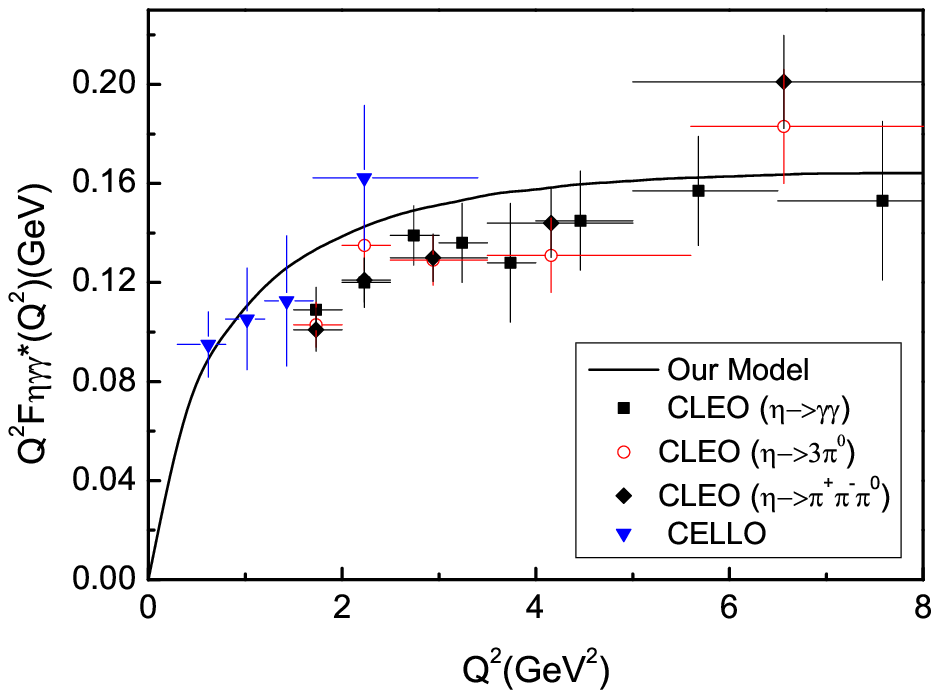}
\caption{The $Q^2$ behavior of the form factor $Q^2
F_{\eta\rightarrow\gamma\gamma^*}(Q^2)$ compared with experimental
data~\cite{Behrend91,Gronberg98}.}
\end{figure}

\begin{figure}[h]
\centering
\includegraphics[0,0][300,250]{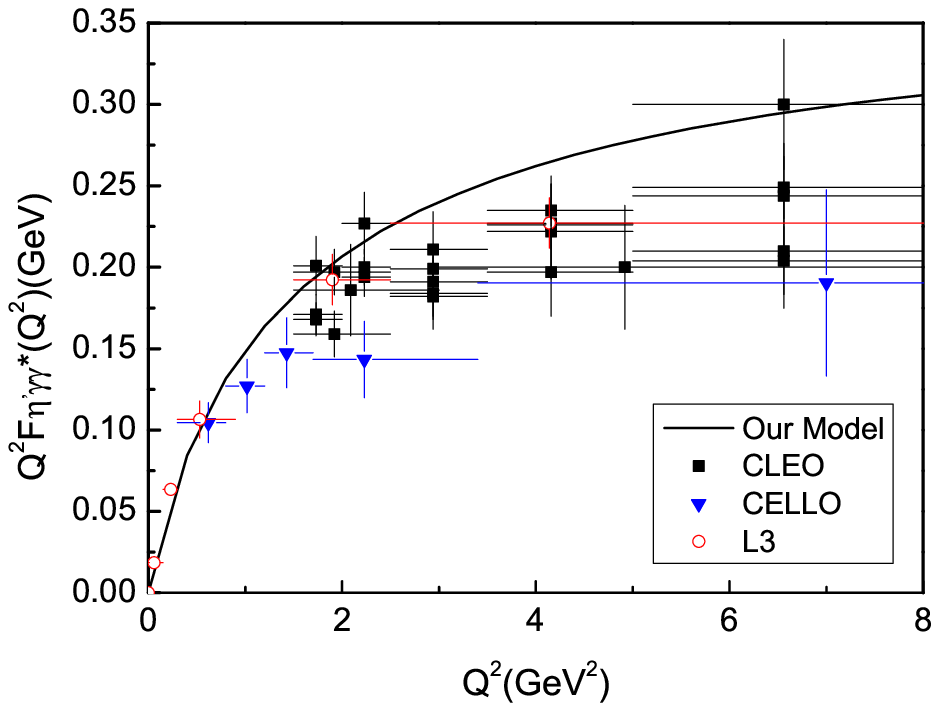}
\caption{The $Q^2$ behavior of the form factor $Q^2
F_{\eta'\rightarrow\gamma\gamma^*}(Q^2)$ compared with experimental
data~\cite{Behrend91,Gronberg98,Acciarri98}.}
\end{figure}

\begin{figure}[h]
\centering
\includegraphics[0,0][300,250]{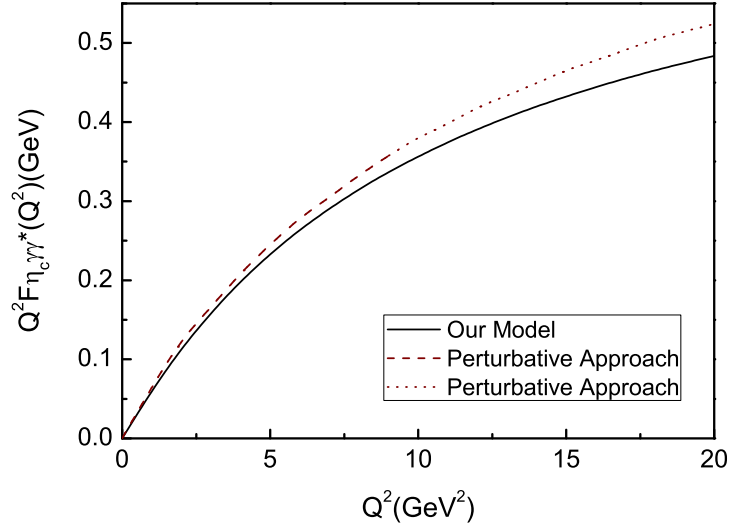}
\caption{Prediction of the $Q^2$ behavior of the form factor $Q^2
F_{\eta_c\rightarrow\gamma\gamma^*}(Q^2)$, compared with the predictions
in the leading order of the perturbative approach~\cite{Feldmann97b}.
The dotted curve of the perturbative approach indicates the $Q^2$ region where
QCD corrections may alter the predictions slightly.}
\end{figure}

\begin{figure}[h]
\centering
\includegraphics[0,0][300,250]{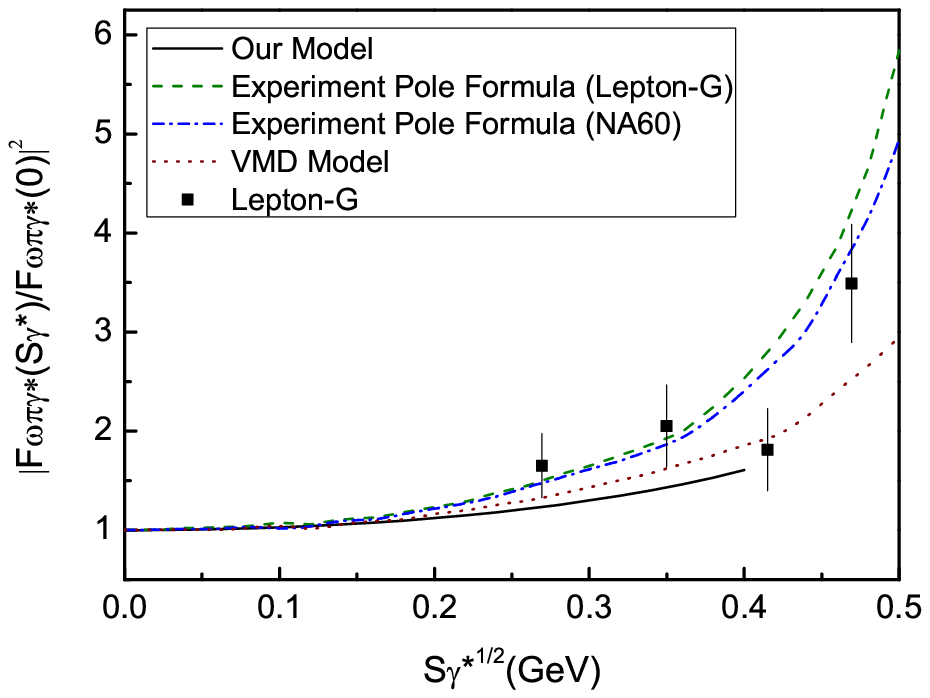}
\caption{The $Q^2$ behavior of the form factor $Q^2
F_{\omega\rightarrow\pi\gamma^*}(Q^2)$ compared with experimental
data~\cite{Arnaldia09,Landsberg85}.}
\end{figure}

\begin{figure}[h]
\centering
\includegraphics[0,0][300,250]{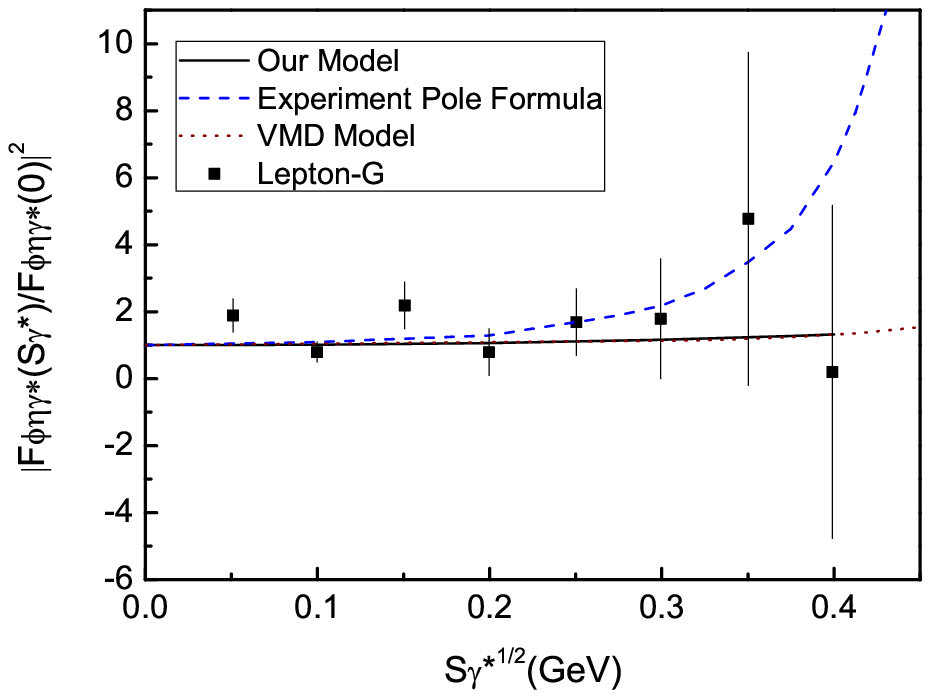}
\caption{The $Q^2$ behavior of the form factor $Q^2
F_{\phi\rightarrow\eta\gamma^*}(Q^2)$ compared with experimental
data~\cite{Achasov01}.}
\end{figure}

\begin{figure}[h]
\centering
\includegraphics[0,0][300,250]{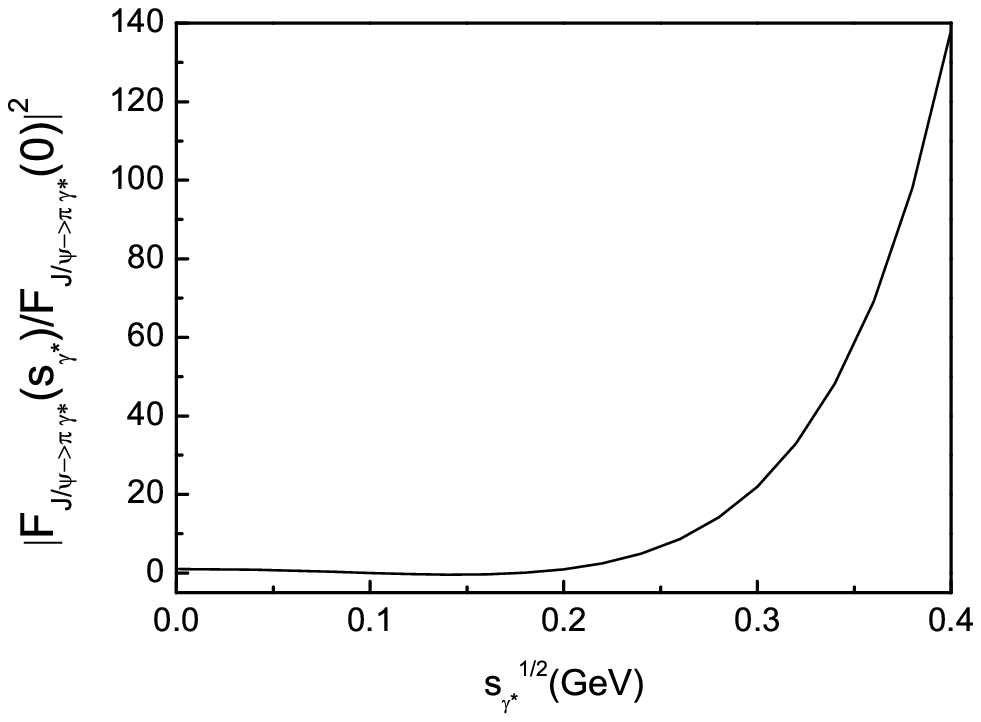}
\caption{Prediction of the $Q^2$ behavior of the form factor $Q^2
F_{J/\psi\rightarrow\pi\gamma^*}(Q^2)$}
\end{figure}

\begin{figure}[h]
\centering
\includegraphics[0,0][300,250]{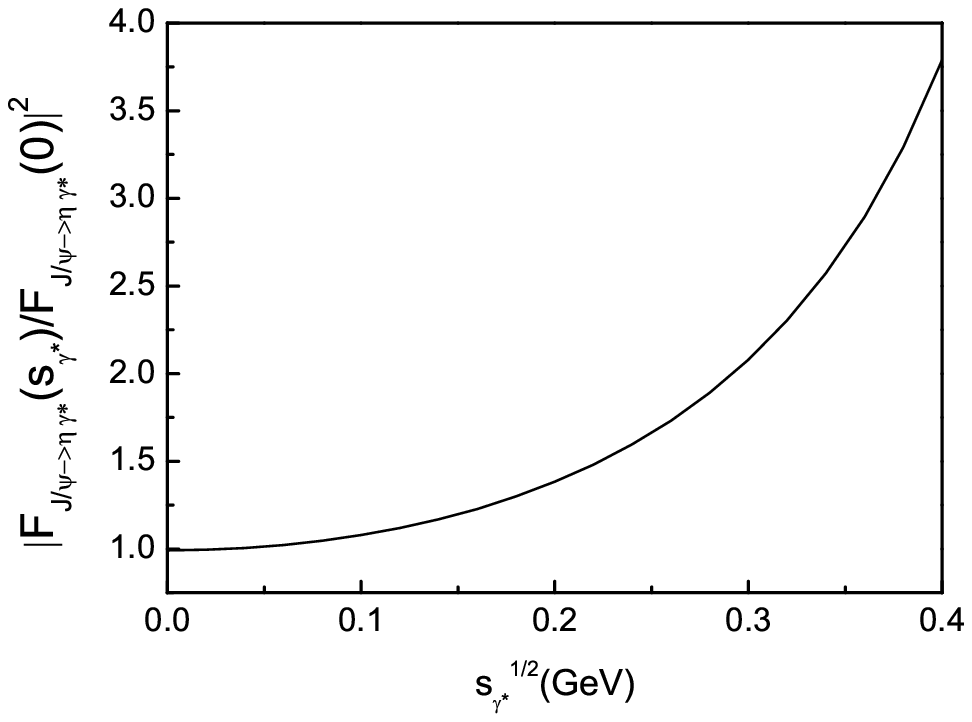}
\caption{Prediction of the $Q^2$ behavior of the form factor $Q^2
F_{J/\psi\rightarrow\eta\gamma^*}(Q^2)$}
\end{figure}

\begin{figure}[h]
\centering
\includegraphics[0,0][300,250]{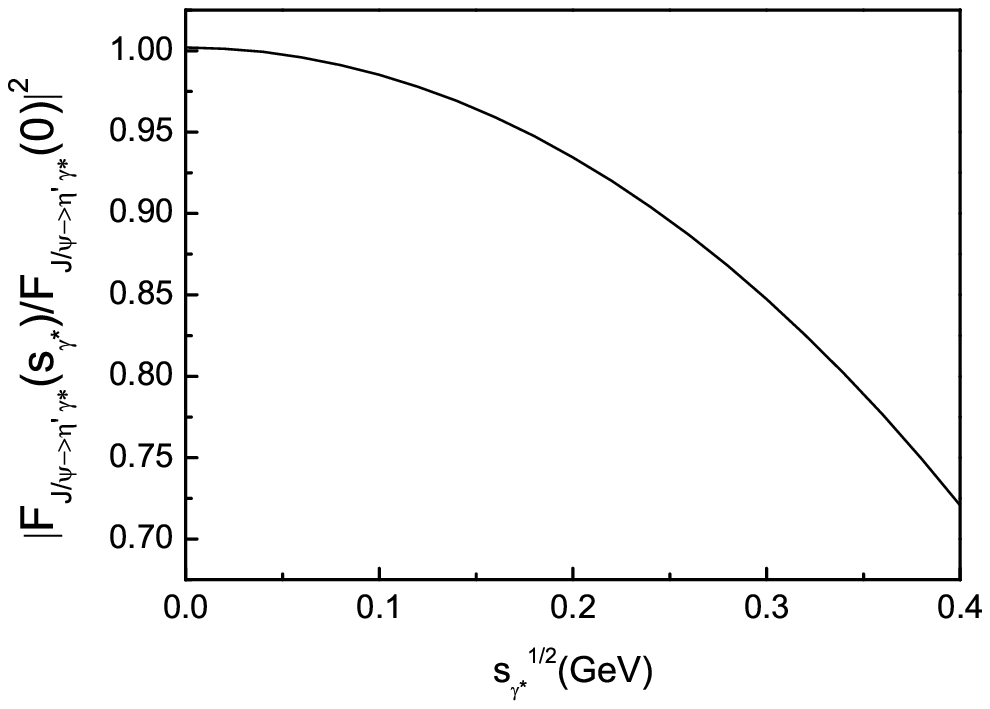}
\caption{Prediction of the $Q^2$ behavior of the form factor $Q^2
F_{J/\psi\rightarrow\eta'\gamma^*}(Q^2)$}
\end{figure}

\begin{figure}[h]
\centering
\includegraphics[0,0][300,250]{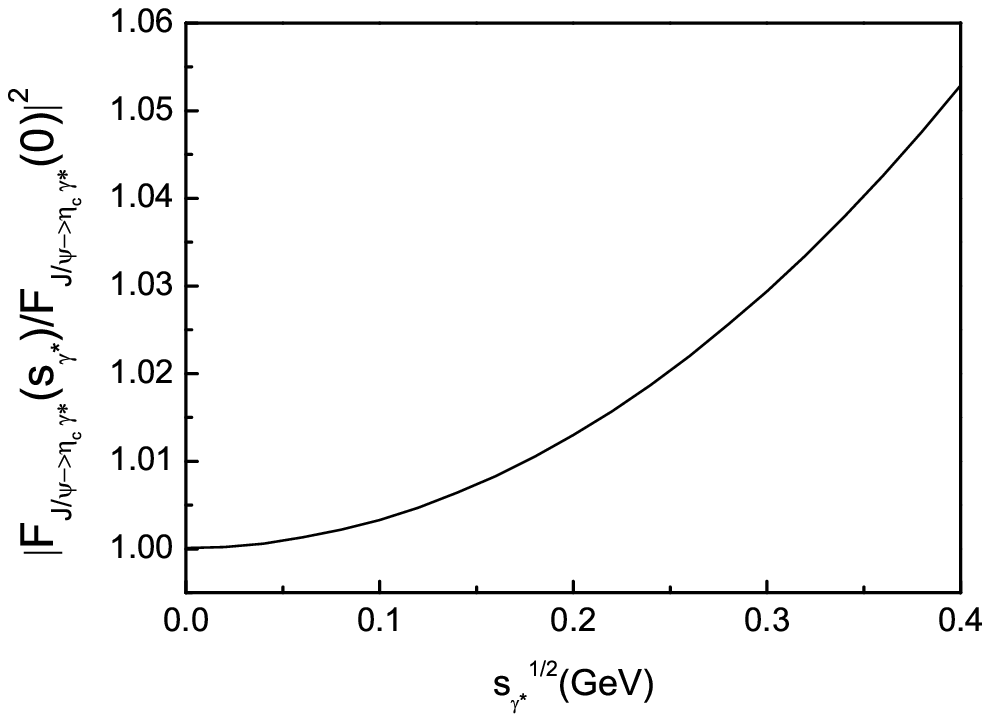}
\caption{Prediction of the $Q^2$ behavior of the form factor $Q^2
F_{J/\psi\rightarrow\eta_c\gamma^*}(Q^2)$}
\end{figure}

We can further use our results to learn the properties of the
intrinsic $c\bar{c}$ component in the light pseudoscalar mesons.
With the $F_{P \rightarrow \gamma \gamma^*}(0)$ (where $P=\pi$,
$\eta$, $\eta'$, $\eta_c$) in Table~\ref{fitting results} and the
mixing matrix $M_s$, we obtain the transition form factors of
unmixed mesons $F_{P_I \rightarrow \gamma \gamma^*}(0)$ (where
$P_I=\pi_I$, $\eta_q$, $\eta_s$, $\eta_{c0}$). Taking them into
Eq.~(\ref{eq:limitting}), we have the values of $f_{\pi}$, $f_{q}$, $f_{s}$, and $f_{c}$:

\begin{eqnarray}
\begin{array}{c}
  f_\pi = 0.0984 \ \mathrm{GeV}, \\
  f_q =  0.0976 \ \mathrm{GeV},\\
  f_s = 0.1298\ \mathrm{GeV},\\
  f_{c} = 0.4874\ \mathrm{GeV}.
\end{array}
\end{eqnarray}

Then we have~\cite{Qian09}

\begin{eqnarray}
\
\left(
  \begin{array}{cccc}
    f_\pi^{\pi} & f_\pi^q & f_\pi^s & f_\pi^c\\
    f_\eta^{\pi} & f_\eta^q & f_\eta^s & f_\eta^c\\
    f_{\eta'}^{\pi} & f_{\eta'}^q & f_{\eta'}^s & f_{\eta'}^c\\
    f_{\eta_c}^{\pi} & f_{\eta_c}^q & f_{\eta_c}^s & f_{\eta_c}^c
  \end{array}
\right) %
&=& M_s
\left(
  \begin{array}{cccc}
    f_\pi & 0 & 0 & 0\\
    0 & f_q & 0 & 0\\
    0 & 0 & f_s & 0\\
    0 & 0 & 0 & f_c
  \end{array}
\right) \nonumber\\
 &=&
\left(
  \begin{array}{cccc}
    0.0974  &  0.0054  & -0.0145  &  0.0168\\
    -0.0106  &  0.0798 &  -0.0729  & -0.0127\\
    0.0058  &  0.0556  &  0.1059  &  0.0219\\
   -0.0039  & -0.0006  & -0.0062  &  0.4855
  \end{array}
\right).
\end{eqnarray}
We see that $f_{\eta'}^c=0.0219\ \mathrm{GeV}=21.9\ \mathrm{MeV}$.
It is compared with previous results in Table~\ref{fetapc}.
$f_{\eta'}^c$ could be considered as the reflection of the intrinsic
charm content of the $\eta'$ meson~\cite{Cao98}, and we see from
Table~\ref{fetapc} that our result of $f_{\eta'}^c$ is in the
similar region with most of the previous results.

\begin{table}[h]
\caption{\label{fetapc}The value of $f_{\eta'}^c$~(MeV) in different models.}
\begin{tabular}{cccccc}
\hline
\hline
{\tabincell{c} {Our\\model}} & { \tabincell{c} {Feldmann\\and Kroll~\cite{Feldmann97a}}} & { \tabincell{c} {Halperin and\\Zhitnitsky~\cite{Halperin97}}} & { \tabincell{c} {Cheng and\\Tseng~\cite{Cheng97}}}& { \tabincell{c} {Cao, Cao, Huang\\and Ma~\cite{Cao98}}} & { \tabincell{c} {Yuan and\\Chao~\cite{Yuan97}}}  \\
\hline
21.9 &-65$-$15 & 50 $-$ 180 & -50 & Around -15 & 40  \\
\hline
\hline
\end{tabular}
\end{table}

In summary, we use the light-cone constituent quark
model to study the tetramixing of pseudoscalar mesons
$\pi$-$\eta$-$\eta'$-$\eta_c$ and vector mesons
$\omega$-$\rho$-$\phi$-$J/\psi$. The parameters of mixing matrices
and meson parameters are determined by fitting our theoretical model
results of the meson decay constants and transition form factors (at
$Q^2=0$) to the experimental data. We also calculate the $Q^2$
behaviors of the meson transition form factors, and these results
are generally in agreement with the experimental data or results
from other models. Our results of the $Q^2$ behaviors of transition
from factors of $J/\psi$ decaying into pseudoscalar mesons could be
regarded as the predictions of our model, as there are no
experimental data at present. The introduction of light quark
components in $J/\psi$ and $\eta_c$ not only allows them to decay
into the light mesons directly without intermediate gluons or
virtual photon but is also helpful for us to understand the
structures of charmonium states better. Considering that the mixing
introduces a $c\bar{c}$ component into the light mesons, and such a
$c\bar{c}$ component is intrinsic to the wave functions and exists
over a time scale independent of any probe momentum, we could
naturally interpret it as the intrinsic charm of these mesons. Our
result of the intrinsic charm content of the $\eta'$ meson
$f_{\eta'}^c$ is also comparable with predictions from other models.

\section*{Acknowledgments}
This work is supported by National Natural Science Foundation of
China (Grants No. 10721063, No. 10975003, and No. 11035003).

\appendix

\section{\label{app:a}}

The SO(4) group elements can be written in terms of the SO(3) group generators~($A_k$ and $B_k$):
\begin{eqnarray}
M&=&R_+R_-,\\
R_+&=&e^{-i\theta_k A_k},~~R_-=e^{-i\theta_{k+2}B_k},~~(k=1,2,3).
\end{eqnarray}
The generators $A_k$ and $B_k$ obey the commuting relations of SO(3) generators~\cite{Pauli65}:
\begin{eqnarray}
[A_i,A_j]=i\varepsilon_{ijk}A_k,~~[B_i,B_j]=i\varepsilon_{ijk}B_k,~~[A_i,B_j]=0.
\end{eqnarray}
We see that the groups have the relation $SO(4)=SO(3)\otimes SO(3)$,
and the generators $A_k$ (as well as $B_k$) (k=1,2,3) could be seen
as the angular momentum operators in each of the three directions.

One form of $A_k$ and $B_k$ is~\cite{Ma04}
\begin{eqnarray}
A_1=\frac{i}{2}\left( \begin{array}{cccc}
0 & -1 & 0 & 0\\
1 & 0 & 0 & 0\\
0 & 0 & 0 & -1\\
0 & 0 & 1 & 0\\
\end{array}\right),
~~A_2=\frac{i}{2}\left( \begin{array}{cccc}
0 & 0 & -1 & 0\\
0 & 0 & 0 & 1\\
1 & 0 & 0 & 0\\
0 & -1 & 0 & 0\\
\end{array}\right),
~~A_3=\frac{i}{2}\left( \begin{array}{cccc}
0 & 0 & 0 & -1\\
0 & 0 & -1 & 0\\
0 & 1 & 0 & 0\\
1 & 0 & 0 & 0\\
\end{array}\right),\\
B_1=\frac{i}{2}\left( \begin{array}{cccc}
0 & 1 & 0 & 0\\
-1 & 0 & 0 & 0\\
0 & 0 & 0 & -1\\
0 & 0 & 1 & 0\\
\end{array}\right),
~~B_2=\frac{i}{2}\left( \begin{array}{cccc}
0 & 0 & 1 & 0\\
0 & 0 & 0 & 1\\
-1 & 0 & 0 & 0\\
0 & -1 & 0 & 0\\
\end{array}\right),
~~B_3=\frac{i}{2}\left( \begin{array}{cccc}
0 & 0 & 0 & 1\\
0 & 0 & -1 & 0\\
0 & 1 & 0 & 0\\
-1 & 0 & 0 & 0\\
\end{array}\right).
\end{eqnarray}
Then
\begin{eqnarray}
R_+=e^{-i\theta_k A_k}=e^{-i\alpha \mathbf{n}\cdot \mathbf{A}}=e^{-i\alpha A_n},
\label{eq:rpappendix}
\end{eqnarray}
where
\begin{equation}
\alpha=\sqrt{\theta_1^2+\theta_2^2+\theta_3^2},
~~\mathbf{n}=\frac{1}{\alpha}(\theta_1,\theta_2,\theta_3).
\end{equation}
From the matrix form of $A_k$, we have
\begin{equation}
A_n=\frac{\theta_k A_k}{\alpha}
=\frac{i}{2\alpha}\left( \begin{array}{cccc}
0 & -\theta_1 & -\theta_2 & -\theta_3\\
\theta_1 & 0 & -\theta_3 & \theta_2\\
\theta_2 & \theta_3 & 0 & -\theta_1\\
\theta_3 & -\theta_2 & \theta_1 & 0\\
\end{array}\right).
\label{eq:an}
\end{equation}
$A_n$ is the angular momentum component of the direction
$\mathbf{n}$. In fact, the matrix form of $A_n$ in Eq.~(\ref{eq:an})
can be diagonalized as
\begin{equation}
A_n'=\left( \begin{array}{cccc}
-\frac{1}{2} & 0 & 0 & 0\\
0 & -\frac{1}{2} & 0 & 0\\
0 & 0 & \frac{1}{2} & 0\\
0 & 0 & 0 & \frac{1}{2}\\
\end{array}\right),
\end{equation}
which is the expression of angular momentum operator $A_n$ in its
eigenstate representation, with the eigenvalues of $A_n$ being
($-1/2$, $-1/2$, $1/2$, $1/2$). The two matrix forms are related as
$A_n' =S^{\dagger} A_n S$, with the transformation matrix
\begin{equation}
S=\frac{1}{\sqrt{2}\alpha}
\left( \begin{array}{cccc}
\frac{\theta_1 \theta_2 +i\theta_3 \alpha}{\sqrt{\theta_2^2+\theta_3^2}} & i\theta_2
& \frac{\theta_1 \theta_2 -i\theta_3 \alpha}{\sqrt{\theta_2^2+\theta_3^2}} & -i\theta_2\\
\frac{\theta_1 \theta_3 -i\theta_2 \alpha}{\sqrt{\theta_2^2+\theta_3^2}} & i\theta_3
& \frac{\theta_1 \theta_3 +i\theta_2 \alpha}{\sqrt{\theta_2^2+\theta_3^2}}  & -i\theta_3\\
0 & \alpha & 0 & \alpha\\
\sqrt{\theta_2^2+\theta_3^2} & -i\theta_1 & \sqrt{\theta_2^2+\theta_3^2} & i\theta_1\\
\end{array}\right).
\end{equation}

In the eigenstate representation of angular momentum $A_n$, the matrix form of $R_+$ is
\begin{equation}
R_+'=e^{-i\alpha A_n}=\left( \begin{array}{cccc}
e^{\frac{i\alpha}{2}} & 0 & 0 & 0\\
0 & e^{\frac{i\alpha}{2}} & 0 & 0\\
0 & 0 & e^{\frac{-i\alpha}{2}} & 0\\
0 & 0 & 0 & e^{\frac{-i\alpha}{2}}\\
\end{array}\right),
\end{equation}
and then we have
\begin{eqnarray}
R_+=SR_+'S^{\dagger}
=\left( \begin{array}{cccc}
\cos \frac{\alpha}{2} & -\frac{\theta_1}{\alpha}\sin \frac{\alpha}{2} & -\frac{\theta_2}{\alpha}\sin \frac{\alpha}{2}
& -\frac{\theta_3}{\alpha}\sin \frac{\alpha}{2}\\
\frac{\theta_1}{\alpha}\sin \frac{\alpha}{2} & \cos \frac{\alpha}{2} & -\frac{\theta_3}{\alpha}\sin \frac{\alpha}{2}
& \frac{\theta_2}{\alpha}\sin \frac{\alpha}{2}\\
\frac{\theta_2}{\alpha}\sin \frac{\alpha}{2} & \frac{\theta_3}{\alpha}\sin \frac{\alpha}{2} & \cos \frac{\alpha}{2}
& -\frac{\theta_1}{\alpha}\sin \frac{\alpha}{2}\\
\frac{\theta_3}{\alpha}\sin \frac{\alpha}{2} & -\frac{\theta_2}{\alpha}\sin \frac{\alpha}{2} & \frac{\theta_1}{\alpha}\sin \frac{\alpha}{2} & \cos \frac{\alpha}{2}\\
\end{array}\right),
\end{eqnarray}
which is the expression of $R_+$ given in Eq.~(\ref{eq:rp}).
Following the same procedure, we also obtain the expression of $R_-$
given in Eq.~(\ref{eq:rm}).

\section{\label{app:b}}

During the numerical calculation, we write $M$ in a more compact
form with eight real parameters~($a$, $b$, $c$, $d$, $p$, $q$, $r$,
and $s$) under the constraints $a^2+b^2+c^2+d^2=1$ and
$p^2+q^2+r^2+s^2=1$~\cite{vanElfrinkhof1897}:
\begin{eqnarray}
M &=&
\left( \begin{array}{cccc}
a & -b & -c & -d\\
b & a & -d & c\\
c & d & a & -b\\
d & -c & b & a\\
\end{array}\right)
\left( \begin{array}{cccc}
p & -q & -r & -s\\
q & p & s & -r\\
r & -s & p & q\\
s & r & -q & p\\
\end{array}\right)\\
&=&\left( \begin{array}{cccc}
\scriptstyle{a p-b q-c r-d s} &  \scriptstyle{-a q-b p+c s-d r}
&  \scriptstyle{-a r-b s-c p+d q} &  \scriptstyle{-a s+b r-c q-d p}\\
\scriptstyle{b p+a q-d r+c s} &  \scriptstyle{-b q+a p+d s+c r}
&  \scriptstyle{-b r+a s-d p-c q} &  \scriptstyle{-b s-a r-d q+c p}\\
\scriptstyle{c p+d q+a r-b s} &  \scriptstyle{-c q+d p-a s-b r}
&  \scriptstyle{-c r+d s+a p+b q} &  \scriptstyle{-c s-d r+a q-b p}\\
\scriptstyle{d p-c q+b r+a s} &  \scriptstyle{-d q-c p-b s+a r}
&  \scriptstyle{-d r-c s+b p-a q} &  \scriptstyle{-d s+c r+b q+a p}\\
\end{array}\right).
\end{eqnarray}
These parameters are related to the six rotation angles as
\begin{eqnarray}
a&=&\cos \frac{\alpha}{2},~~b=\frac{\theta_1}{\alpha}\sin \frac{\alpha}{2},
~~c=\frac{\theta_2}{\alpha}\sin \frac{\alpha}{2},~~d=\frac{\theta_3}{\alpha}\sin \frac{\alpha}{2},\\
p&=&\cos \frac{\beta}{2},~~q=-\frac{\theta_4}{\beta}\sin \frac{\beta}{2},
~~r=-\frac{\theta_5}{\beta}\sin \frac{\beta}{2},~~s=-\frac{\theta_6}{\beta}\sin \frac{\beta}{2}.
\end{eqnarray}
When referring to the mixing of specific types of mesons, the parameters ($a$, $b$, ...)
change to ($a_v$, $b_v$, ...) or ($a_s$, $b_s$, ...) correspondingly.

\end{document}